\documentclass[%
  reprint,
  twocolumn,
 amsmath,amssymb,
 aps,
]{revtex4-1}
\usepackage{graphicx}% Include figure files
\usepackage{dcolumn}% Align table columns on decimal point
\usepackage{bm}% bold math
\usepackage{color}
\usepackage{footnote}
\begin{document}

%----------------------------------------------------------------------------------------
%	TITLE SECTION
%----------------------------------------------------------------------------------------

%\setlength{\droptitle}{-4\baselineskip} % Move the title up
%
%\pretitle{\begin{center}\Huge\bfseries} % Article title formatting
%\posttitle{\end{center}} % Article title closing formatting
\title{ %\Large{
Probing extra Yukawa couplings by precision measurements of Higgs properties
} %} % Article title
\author{Wei-Shu Hou and %}
%\author{
Mariko Kikuchi}
%\email{aaa}
%\affil[1]
\affiliation{Department of Physics, National Taiwan University, Taipei 10617, Taiwan
}

\date{\today} % Leave empty to omit a date
\begin{abstract}
If one removes any \emph{ad hoc} symmetry assumptions,
the general two Higgs doublet model should have additional Yukawa interactions
independent from fermion mass generation, in general involving
flavor changing neutral Higgs couplings.
These extra couplings can affect the discovered
Higgs boson $h$ through fermion loop contributions.
We calculate the renormalized $hZZ$ coupling at one-loop level
%by on-shell and minimal subtraction scheme,
and evaluate the dependence on heavy Higgs boson mass
and extra Yukawa coupling $\rho_{tt}$.
Precision measurements at future colliders can explore
the parameter space, and can give stronger bound on $\rho_{tt}$
than the current bound from flavor experiments.
As a side result, we find that if $\rho_{tt}\cos\gamma < 0$,
where $\cos\gamma$ is the exotic Higgs component of $h$,
the $\rho_{tt}$-induced top loop contribution cancels against
bosonic loop contributions, and one may have alignment without decoupling,
namely $\sin(-\gamma) \simeq 1$, but exotic scalar bosons
could have masses at several hundred GeV.

%11.15.Ex	%Spontaneous breaking of gauge symmetries
%12.15.Ff	%Quark and lepton masses and mixing (see also 14.60.Pq Neutrino mass and mixing)
%14.65.Jk % Other quarks (e.g., 4th generations)
%11.30.Er % Charge conjugation, parity, time reversal, and other discrete symmetries
%14.80.-j	%Other particles (including hypothetical)
% You may use the \texttt{description} environment to structure your abstract; use the optional argument of the \verb+\item+ command to give the category of each item.
%\end{description}
\end{abstract}
%}

%----------------------------------------------------------------------------------------

%\begin{document}


% Print the title
\maketitle

%----------------------------------------------------------------------------------------
%	ARTICLE CONTENTS
%----------------------------------------------------------------------------------------

\section{Introduction}\label{sec:Intro}
The LHC has firmly established the 125 GeV Higgs boson ($h$),
and all data so far are consistent~\cite{LHC_Run1_mass, LHC_Run1_Higgs}
with the predictions of the Standard Model (SM).
But, within measurement errors,
this certainly does not mean that the Higgs sector must be minimal within SM.
There is no theoretical principle that requires the Higgs sector
to be composed of only one weak isodoublet, and
it may well be extended beyond the minimal.

With the existence of one doublet established,
the two Higgs doublet model (2HDM)
is one of the simplest and most reasonable extensions of the Higgs sector,
and often appears in beyond SM new physics models, such as supersymmetry (SUSY).
There are various types of 2HDMs,
the most popular are those with a softly broken $Z_2^{}$ symmetry~\cite{2HDMs_Z2},
which forbids flavor changing neutral Higgs (FCNH) couplings.
The so-called 2HDM II, where each charge type of quarks receive mass
from their own separate Higgs doublet, automatically arises with SUSY.
In part because of this, theoretical and phenomenological properties of
2HDMs with $Z_2^{}$ symmetry have been studied
extensively in the literature~\cite{Branco_2HDM}.
However, since the $Z_2^{}$ symmetry is \emph{ad hoc},
the Yukawa matrices may become too restrictive ``artificially''. % due to the \emph{ad hoc} symmetry.
In the LHC era, the additional Yukawa interactions
should not be determined by such \emph{ad hoc} symmetries,
but by experiments in a bottom-up approach.
After all, so far there is no indication of SUSY at the LHC.

If the Higgs sector is extended to two Higgs doublets, $\Phi$ and $\Phi'$,
there are in general two Yukawa interaction matrices for each type of fermion charge.
As one can always rotate to the basis where only one scalar doublet
develops a vacuum expectation value (VEV),
the Yukawa matrix for the Higgs field with non-zero VEV gives the mass matrix,
hence gets automatically diagonalized,
and these masses and Yukawa couplings are now well measured.
%Since we know values of masses of all fermions, forms of the Yukawa matrices
%have been fixed.
However, the second Yukawa matrix ($\rho_{ij}^f$ with $f=u, d, e$),
i.e. the Yukawa matrix for the scalar field without a VEV,
gives rise to additional Yukawa interactions of
the exotic scalar doublet, which would naturally contain FCNH couplings.
While it was the latter couplings that lead Glashow and Weinberg to impose
discrete symmetries~\cite{2HDMs_Z2} to forbid them,
it was subsequently pointed out that Nature exhibits a
fermion flavor and mass pattern~\cite{Fritzsch:1977vd, Cheng:1987rs}
that may not forbid FCNH couplings involving the third generation~\cite{Hou:1991un},
and 2HDM without $Z_2$ symmetry was called 2HDM III.
We shall just call it the general 2HDM.
Some of the most striking signatures of the scenario are
$t\to ch$~\cite{Hou:1991un, Chen:2013qta} or
$h \to \mu\tau$ decays~\cite{Harnik:2012pb}.

Most components of the second Yukawa matrices have been strongly constrained
by various flavor experiments. However, some components are
still allowed to be $\mathcal{O}(1)$.
For example, the strongest constraint on $\rho_{tt} \equiv \rho_{33}^{u}$
is given by $\bar{B}_{d,s}^0-B_{d,s}^0$ mixing,
%data of at the BABAR
%~\cite{B-B_mixing}
but $\rho_{tt} \sim 1$ is allowed~\cite{AHKKM}.
In this paper, we do not address FCNH couplings, but would like to
suggest indirect detection of the additional Yukawa interactions
via precision measurements of Higgs boson $h$ couplings at future colliders.
The effect of additional Yukawa interactions such as $\rho_{tt}$
appears as deviations in Higgs boson couplings from SM prediction.
Measurement accuracies will be dramatically improved in the future,
first at the high luminosity LHC (HL-LHC), and subsequently
at the International Linear Collider (ILC).
For example, an expected uncertainty (1$\sigma$) of the $hZZ$ coupling
is $\mathcal{O}(1\%)$~\cite{HWGR,HL-LHC} and $\mathcal{O}(0.1\%)$~\cite{HWGR}
at the HL-LHC and ILC, respectively.
Such precision measurements can probe coupling deviations due to
the extra Yukawa interactions.

We calculate the renormalized $hZZ$ coupling at the one-loop level
in the on-shell and minimal subtraction scheme.
Although the one-loop correction to Higgs boson couplings have been
well studied in the 2HDMs with $Z_2^{}$ symmetry~\cite{
KOSY, KKY_2HDM_yukawa, KKY_2HDM_full, Arhrib_IDM, Krause_2HDM, KKS_IDM, Krause:2016xku, Arhrib:2016snv},
such is not the case for the general 2HDM.
We evaluate not only fermion loop contributions,
but also scalar and vector boson loop contributions.
This paper focuses on the top quark loop contributions
to the $hZZ$ couplings as a simple first step.
We evaluate numerically the dependence of $hZZ$ coupling on
the heavy Higgs boson mass and additional Yukawa coupling parameter $\rho_{tt}$,
and elucidate ``alignment without decoupling''~\cite{Gunion:2002zf, Pilaftsis, Craig:2013hca}
that the general 2HDM could harbor.
That is, when the top loop contribution cancels against
the bosonic loop contributions, one could have alignment
($h$ is close to SM Higgs) without pushing the extra Higgs bosons
to become superheavy.
We illustrate what parameter space in $\rho_{tt}$
the HL-LHC and ILC precision measurements can explore for
several heavy Higgs boson masses.
We discuss whether the precision measurements can give
stronger bound on $\rho_{tt}$ than the current bound from
${B}_{d}^0$ and $B_{s}^0$ mixings,
and the constraint from future prospects for new scalar boson search at the LHC.

This paper is organized as follows.
In Sec. II and III, we briefly review the tree level properties of
the 2HDM Higgs potential and the Yukawa interaction, respectively,
to fix notation and motivate our study.
We present our calculational scheme in Sec. IV for one-loop corrections
to the Higgs boson couplings in the general 2HDM.
In Sec. V, we numerically study the deviation in $hZZ$ coupling
as a function of $\rho_{tt}$ and extra scalar boson masses,
as well as dependence on Higgs mixing,
and then compare with future precision measurement sensitivities.
Conclusion is given in Sec. VI, while
various formulae are collected in an Appendix.

%------------------------------------------------

%\section{Model}
\section{Higgs potential}\label{sec:Pote}
The Higgs potential of the general two Higgs doublet model (2HDM) is given by
   \begin{align}
        V &= \mu_{11}^2 |\Phi|^2 +\mu_{22}^2 |\Phi'|^2 - (\mu_{12}^2 \Phi^\dagger \Phi' + h.c.) \notag\\
   &+\frac{\eta_1}{2}|\Phi|^4 + \frac{\eta_2}{2}|\Phi'|^4 + \eta_3|\Phi|^2|\Phi'|^2
   + \eta_4(\Phi^\dagger \Phi')(\Phi'^\dagger \Phi^{}) \notag\\
   &+\left\{\frac{\eta_5^{}}{2}(\Phi^\dagger \Phi')^2
   + (\eta_6^{}|\Phi|^2 + \eta_7^{}%\color{black}
      |\Phi'|^2)(\Phi^\dagger \Phi') + h.c.\right\},
   \label{eq:pote2}
   \end{align}
   where $\mu_{12}^2$, $\eta_5$, $\eta_6$ and $\eta_7$ can be complex,
   while the latter two are absent from 2HDM with $Z_2$ symmetries.
The two doublet fields can be parameterized as
   \begin{align}
   \Phi &= \left(\begin{array}{c}
     G^+ \\
     \frac{1}{\sqrt{2}}(\phi_1 + v + i G^0) \\
   \end{array}\right), \notag\\
      \Phi' &= \left(\begin{array}{c}
     H^+ \\
     \frac{1}{\sqrt{2}}(\phi_2 + i A) \\
     \end{array}\right).
     \end{align}
   where, without loss of generality~\cite{Davidson_Haber, Haber_Neil},
   $\Phi$ is taken as the one with non-zero vacuum expectation value
   (VEV), while $\Phi'$ has no VEV.

After imposing the minimization conditions,
$\mu_{11}^2$ and $\mu_{12}^2$ are expressed in terms of other parameters as
\begin{align}
  \mu_{11}^2 = -\frac{\eta_1^{}}{2}v^2, \,\,\,\,
  \mu_{12}^2 = \frac{\eta_6^{}}{2}v^2,
\end{align}
and the mass terms of the Higgs potential become
 \begin{align}
   V_\textrm{mass} &=
     (\phi_1,\; \phi_2)M_\textrm{even}^2\left(\begin{array}{c}
     \phi_1\\
     \phi_2\\
   \end{array}\right)+
   (G^0,\; A)M_\textrm{odd}^2\left(\begin{array}{c}
     G^0\\
     A\\
   \end{array}\right) \notag\\
   &+
   (G^-,\; H^-)M_\pm^2\left(\begin{array}{c}
     G^+\\
     H^+\\
   \end{array}\right),
 \end{align}
where the CP-odd $M_\textrm{odd}^2$ and
the charged $M_{\pm}^2$ matrices are diagonal,
with nonzero eigenvalues given by
 \begin{align}
  m_A^2 & = \mu_{22}^2 +\frac{v^2}{2}(\eta_3 +\eta_4^{} - \eta_{5}^{}), \\
  m_{H^\pm}^2 &= \mu_{22}^2 + \frac{\eta_3^{}}{2}v^2.
 \end{align}
For the CP-even $M_\textrm{even}^2$ matrix, one has
\begin{align}
  M_\textrm{even}^2 =
  \left(\begin{array}{cc}
    \eta_1^{}v^2 & \eta_6^{} v^2 \\
    \eta_6^{} v^2 & \mu_{22}^2 + \frac{v^2}{2} (\eta_3^{} + \eta_4^{} + \eta_5^{})\\
    \end{array}\right),
\end{align}
which is diagonalized by the rotation matrix $R$ with mixing angle $\gamma$,
\begin{align}
  R^T(\gamma) M_\textrm{even}^2 R(\gamma)  =
    \left(\begin{array}{cc}
    m_H^2 & 0 \\
    0 & m_h^2 \\
  \end{array}\right), \notag\\
  \,\,\,\,\, R(\gamma)  =  \left(\begin{array}{cc}
    \cos\gamma & -\sin\gamma \\
    \sin\gamma & \cos\gamma \\
  \end{array}\right),
\end{align}
where we keep the convention of 2HDM II,
and $H$, $h$ are CP-even mass eigenstates.
The mixing angle $\gamma$ is expressed by
\begin{align}
   &\sin2\gamma \notag\\
    =& \frac{2\eta_6^{}v^2}{\sqrt{[v^2(2\eta_1-\eta_3-\eta_4-\eta_5)/2 - \mu_{22}^2 ]^2+ 4(\eta_6 v^2)^2}}\,.
\end{align}
The isospin states $\phi_1$ and $\phi_2$ are related to
the mass eigenstates $H$ and $h$ by
\begin{align}
  \left(\begin{array}{c}
    \phi_1 \\
    \phi_2 \\
    \end{array}\right) = R(\gamma) \left(\begin{array}{c}
    H \\
    h \\
    \end{array}\right), \label{eq:cp-even}
  \end{align}
where $h$ is the 125 GeV boson, and $\sin\gamma \to -1$
corresponds to the SM, or alignment limit.
For the decoupling limit in which the extra Higgs bosons
are much heavier than the electroweak scale, i.e.
$\varepsilon \equiv v^2/m_{H,\, A, \, H^\pm}^2 \ll 1$,
the mixing angle $\gamma$ can be approximated by
\begin{align}
  \cos\gamma \simeq F_\textrm{Sign}\, \varepsilon
  + \mathcal{O}\left(\varepsilon^2\right), \quad {\rm (decoupling)}
  \label{eq:sin_deco}
  \end{align}
where $F_\textrm{Sign}$ is the sign of $\sin\gamma$.

In summary, some parameters in the Higgs potential of Eq.~(\ref{eq:pote2})
are rewritten with physical parameters as
\begin{align}
  \eta_1^{} & = \frac{1}{v^2}\left( m_H^2 c_\gamma^2 + m_h^2 s_\gamma^2 \right),  \label{eq:npara1}\\
  \eta_3^{} & = - \frac{2}{v^2}(\mu_{22}^2 - m_{H^\pm}^2), \label{eq:npara3}\\
  \eta_4^{} & = \frac{1}{2v^2}\left(2 m_A^2 - 4 m_{H^\pm}^2 + m_h^2 + m_H^2 \right. \nonumber \\
   & \quad\quad\quad\quad\quad\quad\quad\quad\quad\ \, \left. +\, (m_h^2 - m_H^2) c_{2\gamma}^{} \right),
     \label{eq:npara4}\\
 \eta_5^{} & = \frac{1}{2v^2}\left(-2 m_A^2 + m_h^2 + m_H^2
           + (m_h^2 - m_H^2) c_{2\gamma}^{}\right), \label{eq:npara5}\\
 \eta_6^{} & = \frac{s_\gamma^{} c_\gamma^{}}{v^2}(-m_h^2 + m_H^2), \label{eq:npara6}
\end{align}
where $s_\gamma = \sin\gamma$ (and likewise for $c_\gamma$, $c_{2\gamma}$),
and
\begin{align}
 \mu_{11}^2 & = - \frac{1}{2} (m_h^2 s_\gamma^2 + m_{H}^2c_\gamma^2), \label{eq:npara8}\\
 \mu_{12}^2 & = -\frac{s_{2\gamma}^{}}{4} (m_h^2 - m_H^2). \label{eq:npara10}
\end{align}
Note that $\eta_2^{}$, $\eta_7$ and $\mu_{22}^2$ remain as free parameters,
as they cannot be expressed in terms of mass and mixing parameters as above.
Altogether, there are 9 independent parameters in the potential.

%% constraints
Dimensionless parameters of the Higgs potential are restricted
by theoretical constraints.
In this paper, we take into account the following constraints:
\begin{itemize}
\item Perturbativity \\
  The perturbative bound requires that all dimensionless parameters be
  smaller than some criterion constants, i.e. $\eta_i \leq \xi_i$, $i=1-7$.
  In all analyses in this paper, we take $\xi_i = 2$.
  While somewhat arbitrary, the point is to keep Higgs parameters
  in perturbative realm.
\item Vacuum stability \\
The vacuum stability bound means the potential should be bounded from below
 in all field directions.
This requires the value of the potential to be positive at large $|\Phi|$ and $|\Phi'|$.
In the analyses of this paper,
we use the vacuum stability condition given in Ref.~\cite{Ivanov}.
  \end{itemize}

%------------------------------------------------
\section{Yukawa interactions}\label{sec:Yukawa}

In this section, we discuss the Yukawa interaction.

\subsection{Exotic Yukawas and the alignment limit}
The general Yukawa Lagrangian for 2HDM is
\begin{align}
     -\mathcal{L}_\textrm{Yukawa}^{} &=
     \bar{Q}_{L,i}^{}(V_{\textrm{CKM}}^\dagger)_{il}\left( \kappa_{u_l u_j}^{} \tilde{\Phi}
       + \rho_{u_l u_j}^{} \tilde{\Phi}'\right)u_{R,j}
     \notag\\
     &+  \bar{Q}_{L,i}^{}\left(\kappa_{d_i d_j}^{} \Phi + \rho_{d_i d_j}^{}\Phi'\right)d_{R,j} \notag\\
     &+ \bar{L}_{L,i}^{}\left(\kappa_{e_i e_j}^{} \Phi + \rho_{e_i e_j}^{}\Phi'\right)e_{R,j}
     + h.c., \label{eq:Yukawa}
\end{align}
with
\begin{align}
     & Q_{L,i} =\left(\begin{array}{c}
       (V_{\textrm{CKM}}^\dagger)_{ij} u_{L,j} \\
       d_{L,i}
  \end{array}\right), \,\,
  L_{L,i} = \left(\begin{array}{c}
       (V_{\textrm{MNS}}^\dagger)_{ij} \nu_{L,j} \\
       e_{L,i}
  \end{array}\right),
\end{align}
where $V_{\textrm{CKM}}$ and $V_{\textrm{MNS}}$ are
the Kobayashi-Maskawa and Maki-Nakagawa-Sakata matrices, respectively.
In Eq.~(\ref{eq:Yukawa}), $\kappa_{f_i f_j}^{}$ multiplied by $v$
corresponds to the mass matrix as $m_{ij}^{} =\delta_{ij} \, \kappa_{f_i f_i}^{} \,v/\sqrt{2}$,
because only $\Phi$ gives VEV.
On the other hand, $\rho_{f_i f_j}^{}$ are the additional Yukawa interactions
of the exotic doublet $\Phi'$, which are in general not diagonal.
Rather than imposing a $Z_2$ symmetry~\cite{2HDMs_Z2} to eliminate
off-diagonal FCNH couplings, the viewpoint promoted here is that they should
be as constrained by data, which is why we call this the general 2HDM.

If $\rho_{f_i f_j}^{}$ are Hermitian matrices, the interaction terms in the mass basis are
\begin{align}
  \mathcal{L}_\textrm{Yukawa}^{}&=  \bar{f}_{i} \lambda_{hf_i f_j} hf_{j}
  + \bar{f}_{i}\lambda_{Hf_i f_j}H f_{j} 
  +i\bar{f}_{i} \lambda_{Af_i f_j} A \gamma_5^{}f_{j} \notag\\
  &-\big[\bar{u}_{i}\left(
  V_{\rm CKM}\rho_{}^dP_R - \rho^{u\dagger}V_{\rm CKM}P_L \right)_{ij}H^+ d_{j} \notag\\
  & +\bar{\nu}_{L,i}\left(
  V_{\rm MNS}\rho_{}^e \right)_{ij}H^+ e_{R,j} +\textrm{h.c.}\big],
\end{align}
where
\begin{align}
  \lambda_{hf_i f_j}^{} &= -
  \frac{m_{f_i}}{v}\sin(-\gamma)\, \delta_{ij}
  - \frac{\rho_{f_if_j}^{}}{\sqrt{2}}\cos\gamma, \label{eq:lff}\\
  %%%
  \lambda_{Hf_i f_j}^{} &= -
  \frac{m_{f_i}}{v}\cos\gamma\, \delta_{ij}
  + \frac{\rho_{f_if_j}^{}}{\sqrt{2}}\sin(-\gamma), \label{eq:hff}\\
  %%%
  \lambda_{Aff}^{} &= \frac{\rho_{f_if_j}^{}}{\sqrt{2}}. \label{eq:aff}
  \end{align}

As all evidence support $h$ to be consistent with the Higgs boson of the SM,
we consider Yukawa coupling constants close to the alignment limit.
That is, we introduce a parameter $x$ defined as
\begin{align}
  \gamma = -\frac{\pi}{2} + x,
  \label{eq:x}
\end{align}
where $x \to 0$ corresponds to the alignment limit.
In this limit,
the coefficients $\lambda_{\phi f_i f_j}^{}$ of the Yukawa interaction vertex
$\phi \bar{f}_i f_j$ can be approximated by
\begin{align}
  \lambda_{hf_i f_j}^{} &= - \frac{m_{f_i}^{}}{v}\,\delta_{ij} - \frac{x}{\sqrt{2}}\,
   \rho_{f_if_j} 
  +\mathcal{O}(x^2), \label{eq:Yuk-h} \\
  \lambda_{Hf_i f_j}^{} &=  \frac{1}{\sqrt{2}}\rho_{f_if_j}^{} 
  - \frac{m_{f_i}^{}}{v}\,\delta_{ij}\,x
  +\mathcal{O}(x^2), \label{eq:Yuk-H} \\
  \lambda_{A f_i f_j}^{} &= \frac{1}{\sqrt{2}}\rho_{f_if_j}^{}\gamma_5^{}
  + \mathcal{O}(x^2). \label{eq:Yuk-A}
\end{align}
While $h$ does pick up a small component of exotic couplings
(including FCNH), in this paper, we shall mostly be interested
in the extra $\rho_{tt}$ coupling of exotic Higgs bosons,
where we have dropped the $u$ quark superscript.

We would like to mention the difference between $x$ and
$\varepsilon$, which is introduced in the previous section.
While $x \to 0$ is the alignment limit,
the limit $\varepsilon \to 0$ means the decoupling of the heavy exotic bosons.
As can be seen from Eq.~(\ref{eq:sin_deco}),
decoupling is a special case of alignment.
%------------------------------------------------

%\section{Experimental constraints}
\subsection{Experimental constraints on $\rho_{tt}$}

%~\footnote{
Elements of $\rho_{ij}^{f}$ for down type quarks and charged leptons
are constrained rather strongly by
%leptonic neutral meson decay
various $B$ meson decay and lepton flavor violation processes~\cite{crivellin}.
%}.
One should, however, keep an eye on $\rho_{23}^e \equiv \rho_{\mu\tau}$,
which can generate $h \to \mu\tau$~\cite{Harnik:2012pb} with
Yukawa coupling strength $\rho_{\mu\tau}\,x$ via Eq.~(\ref{eq:Yuk-h}).
A hint from 8 TeV data by CMS~\cite{Khachatryan:2015kon}
might reappear in the 2016 data set at 13 TeV that is much larger than obtained in 2015.
Similarly, $\rho_{23}^u \equiv \rho_{ct}$ may generate
$t \to ch$ decay~\cite{Hou:1991un, Chen:2013qta},
which is being pursued at the LHC~\cite{Aad:2015pja, Khachatryan:2016atv}.
We note that if these decays are absent, it does not
necessarily imply small $\rho_{\mu\tau}$ and $\rho_{ct}$,
but may reflect the alignment limit of $x \sim 0$.

We are mainly interested in the extra diagonal coupling $\rho_{tt}$
of the exotic Higgs $H$, as the SM Yukawa coupling
$\lambda_t \equiv \kappa_{tt} \simeq 1$ for $h$ is the largest known coupling.
The current bound on $\rho_{tt}$ comes mainly from $B_{d,s}$ mixing and $B \to X_s \gamma$.
It is found~\cite{AHKKM} that the latest $B_{s}$ mixing data gives the 95\% C.L. bound
$|\rho_{tt}| < 1$ ($-0.35 < \rho_{tt} < 0.2$) for $\rho_{ct} = 0$ ($\rho_{ct} = 0.1$),
for charged Higgs boson mass $m_{H^+}^{} = 500$ GeV.
For $m_{H^+}=500$ GeV and $\rho_{ct} = 0$, the region with $\rho_{tt} < -1$ and
$0.6 < \rho_{tt}$ for $\rho_{bb} = \kappa_{bb}$ has been
excluded by data on the $B \to X_s \gamma$ process.
However, if $|\rho_{bb}|$ is less than about 0.005, $\rho_{tt}^{}$
is practically not constrained.
In any case and for our purpose,
if we consider the situation where all components of
$\rho_{ij}^{f}$ matrices are zero except for $\rho_{tt}$,
the strongest bound is $|\rho_{tt}| < 1$~\cite{AHKKM} (for $m_{H^+} = 500$ GeV),
which comes from $B_s$ mixing.
It is intriguing that the second top Yukawa coupling
could be as strong as the SM Higgs boson.

Collider experiments can in principle provide constraints on
$\rho_{tt}^{}$ %and masses of extra scalar bosons
by direct search of the heavy scalar bosons.
Unfortunately, while mass bounds on exotic vector bosons
have been pursued in $t\bar{t}$ resonance searches~\cite{Chatrchyan:2013lca, ATLAS_ttbar_8},
the situation is unclear when it comes to heavy scalars.
This is due to interference with the production of such a boson,
which involves the top quark in the triangle loop
as a consequence of $\rho_{tt} \neq 0$.

The situation for heavy Higgs boson search through
$gg \to S \to t\bar{t}$ process at the LHC has been
assessed recently in Ref.~\cite{carena_zhen},
where the expected 95\% C.L. exclusion limits on the top quark Yukawa
coupling of additional CP-even and CP-odd scalar bosons are evaluated
assuming several LHC scenarios.
Although the simplified model in Ref.~\cite{carena_zhen} is
not the same as the general 2HDM,
for $\cos\gamma \simeq 0$, i.e. $x \simeq 0$,
the results can be applied to the general 2HDM.
For LHC at 13 TeV collision energy and
integrated luminosity of $30$ fb$^{-1}$ (i.e. 2016 data),
one could survey the region of $\rho_{tt}^{} > 2.4$ by using
the $A \to t\bar{t}$ process for $m_A^{} = 500$ GeV.
For $300$ fb$^{-1}$,
the expected bound is improved to $\rho_{tt} > 1.4$ ($\rho_{tt} > 2.6$) for
$m_A^{} = 500$ GeV (1 TeV) using conservative assumptions for
efficiency and systematic uncertainty,
and $\rho_{tt} > 0.5$ ($\rho_{tt} > 0.9$) for $m_A^{} = 500$ GeV (1 TeV)
using more aggressive assumptions.
In the case where $A$ is heavier,
the exclusion limit on $\rho_{tt}$ becomes further relaxed.

It is in part this difficulty of probing $\rho_{tt}$ directly
at LHC via $t\bar t$ scalar resonances
that motivates our indirect, precision measurement approach.

%------------------------------------------------
\section{Renormalization}\label{sec:Reno}

We now discuss renormalization of the scalar sector,
towards the indirect, precision measurement approach.

\subsection{Parameter shift}\label{sec:shift}

As mentioned in Sec.~\ref{sec:Pote}, there are 9 independent parameters,
  \begin{align}
    m_h^2, m_H^2, m_A^2, m_{H^\pm}^2, \gamma, v, \mu_{22}^2, \eta_2^{}, \eta_7^{},
  \end{align}
which get shifted by
  \begin{align}
  &  m_\phi^2 \to m_\phi^2 + \delta m_\phi^2, \,\,\,(\phi=h, H, A,\,
    \textrm{and}\, H^\pm), \\
    & \gamma \to \gamma + \delta \gamma, \,\,\,
    v \to v+ \delta v, \,\,\, \mu_{22}^2 \to \mu_{22}^2 + \delta \mu_{22}^2, \\
    & \eta_2^{} \to \eta_2^{} + \delta \eta_2^{}, \,\,\,
    \eta_7^{} \to \eta_7^{} + \delta \eta_7^{}.
  \end{align}
The CP-even, CP-odd and charged components of the doublet fields
are corrected by
  \begin{align}
    \left(\begin{array}{c}
      \phi_1 \\
      \phi_2 \\
    \end{array}\right) &\to \tilde{Z}_\textrm{even}
    \left(\begin{array}{c}
      \phi_1 \\
      \phi_2 \\
    \end{array}\right), \,\,\,
        \left(\begin{array}{c}
      G^0 \\
      A \\
    \end{array}\right) \to \tilde{Z}_\textrm{odd}
    \left(\begin{array}{c}
      G^0 \\
      A \\
    \end{array}\right), \notag\\
        \left(\begin{array}{c}
      G^\pm \\
      H^\pm \\
    \end{array}\right) &\to \tilde{Z}_\pm
    \left(\begin{array}{c}
      G^\pm \\
      H^\pm \\
    \end{array}\right), \label{eq:field_shift}
    \end{align}
  where $\tilde{Z}_\textrm{even}$, $\tilde{Z}_\textrm{odd}$ and $\tilde{Z}_\pm^{}$ are real $2\times 2$ matrices.
We here define $\tilde{Z}_\textrm{odd}$ and
  $\tilde{Z}_\pm^{}$ as follows,
  \begin{align}
    \tilde{Z}_\textrm{odd}  &= \left(\begin{array}{cc}
      1 + \frac{1}{2}\delta Z_{G0}^{} & \delta C_{GA} \\
      \delta C_{AG} & 1 + \frac{1}{2}\delta Z_A^{} \\ \end{array}\right), \\
    \tilde{Z}_\pm &= \left(\begin{array}{cc}
      1 + \frac{1}{2}\delta Z_{G^\pm}^{} & \delta C_{GH} \\
      \delta C_{HG} & 1 + \frac{1}{2}\delta Z_{H^\pm}^{} \\ \end{array}\right).
  \end{align}

For CP-even states, from Eqs.~(\ref{eq:cp-even}) and (\ref{eq:field_shift}),
the relation between bare mass eigenstates and
renormalized mass eigenstates can be derived as
  \begin{align}
        \left(\begin{array}{c}
      H \\
      h \\
        \end{array}\right)_B& = R(-\gamma)_B \, \tilde{Z}_\textrm{even}
        \left(\begin{array}{c}
      \phi_1 \\
      \phi_2 \\
        \end{array}\right) \notag\\
        &= R(-\delta\gamma) R(-\gamma) \, \tilde{Z}_\textrm{even}R(\gamma)
        \left(\begin{array}{c}
      H \\
      h \\
        \end{array}\right) \notag\\
        &= R(-\delta\gamma) \, Z_\textrm{even}
        \left(\begin{array}{c}
      H \\
      h \\
        \end{array}\right).
  \end{align}
where $Z_\textrm{even}$ is defined as,
  \begin{align}
    Z_\textrm{even} &= \left(\begin{array}{cc}
      1 + \frac{1}{2}\delta Z_H^{} & \delta C_{Hh} \\
      \delta C_{hH} & 1 + \frac{1}{2}\delta Z_h^{} \\ \end{array}\right).
  \end{align}
  Therefore, CP-even mass eigenstates are shifted as
  \begin{align}
    \left(\begin{array}{c}
      H\\
      h\\
    \end{array}\right) \to
    \left(\begin{array}{cc}
      1 + \frac{1}{2}\delta Z_H^{} & \delta C_{Hh}+\delta \gamma \\
      \delta C_{hH}-\delta\gamma & 1 + \frac{1}{2}\delta Z_h^{} \\ \end{array}\right)   \left(\begin{array}{c}
      H\\
      h\\
    \end{array}\right).
 \label{eq:Hh-shift}
  \end{align}
We emphasize that mixing counterterms $\delta C_{XY}^{}$ are
not symmetric, i.e. $\delta C_{Hh}^{} \neq \delta C_{hH}$.

In addition to the above parameters, counterterms of two tadpoles for $\phi_1$ and
$\phi_2$ should be introduced at higher order,
\begin{align}
 T_1 \to T_1 + \delta T_1, \,\,\, T_2 \to T_2 + \delta T_2,
\end{align}
where $T_1$ and $T_2$ on the right-hand sides have to become
zero by minimization conditions of the Higgs potential.
Therefore, the renormalized tadpoles are
\begin{align}
  \hat{T}_1 &= \delta T_1 + T_1^\textrm{1PI}, \\
  \hat{T}_2 &= \delta T_2 + T_2^\textrm{1PI},
  \end{align}
where $T_i^\textrm{1PI}$ are the one particle irreducible (1PI) diagram
contributions to the tadpole of $\phi_i$.
Explicit forms of their fermion loop contributions are given in the Appendix.

\subsection{Renormalized two-point functions}
The renormalized two point functions $\hat{\Pi}_{XY}^{}$ are expressed as
\begin{align}
  \hat{\Pi}_{hh}^{} [p^2] &=
  (p^2-m_h^2)\delta Z_h -\delta m_h^2 + \frac{s_\gamma^2}{v}\delta T_1^{}\notag\\
  &-\frac{2s_\gamma c_\gamma}{v}\delta T_2^{} + \Pi_{hh}^{1\textrm{PI}}[p^2],
  \label{eq:pi_hh}\\
  \hat{\Pi}_{HH}^{} [p^2] &=
  (p^2-m_H^2)\delta Z_H -\delta m_H^2 + \frac{c_\gamma^2}{v}\delta T_1^{}
  \notag\\ &+\frac{2s_\gamma c_\gamma}{v}\delta T_2^{} + \Pi_{HH}^{1\textrm{PI}}[p^2],
  \label{eq:pi_HH}\\
  \hat{\Pi}_{AA}^{}[p^2] &=
  (p^2 - m_A^2)\delta m_A^2 -\delta m_A^2 + \Pi_{AA}^{1\textrm{PI}}[p^2],
  \label{eq:pi_AA}\\
  \hat{\Pi}_{H^+H^-}^{}[p^2]&=
  (p^2 - m_{H^\pm}^2)\delta Z_{H^\pm}^{} - \delta m_{H^\pm}^2 +\Pi_{H^+H^-}^{1\textrm{PI}}[p^2]. \label{eq:pi_HpHm}
  \end{align}
Renormalized scalar mixing effects are given by
  \begin{align}
  \hat{\Pi}_{hH}^{} [p^2]&=
  p^2(\delta C_{hH}^{} + \delta C_{Hh}^{}) + m_h^2(\delta \gamma -\delta C_{hH}^{})
  \notag\\
  &- m_H^2(\delta \gamma + \delta C_{Hh}) \notag\\
  &- \frac{s_\gamma c_\gamma}{v}\delta T_1^{}
  +\frac{c_{2\gamma}^{}}{v}\delta T_2^{} + \Pi_{hH}^{1\textrm{PI}}[p^2], \\
  \hat{\Pi}_{AG}^{}[p^2]&=
  p^2(\delta C_{AG} + \delta C_{GA}) -m_A^2 \delta C_{AG}^{}
  + \frac{1}{v}\delta T_2^{} \notag\\ &+ \Pi_{AG}^{1\textrm{PI}}[p^2], \\
  \hat{\Pi}_{HG}^{}[p^2]&=
  p^2(\delta C_{HG} + \delta C_{GH}) -m_{H^\pm}^2 \delta C_{HG}^{}
  + \frac{1}{v}\delta T_2^{} \notag\\ &+ \Pi_{HG}^{1\textrm{PI}}[p^2].
  \end{align}

  \subsection{Renormalization conditions}
In this subsection, we discuss how the counterterms can be determined by
the renormalization conditions.

We determine the counterterms of tadpoles by the following conditions,
\begin{align}
  \hat{T}_h =0, \,\,\, \hat{T}_H=0,
\end{align}
hence
\begin{align}
  \delta T_h^{} = -T_h^\textrm{1PI}, \,\,\,
  \delta T_H^{} = -T_H^\textrm{1PI},
\end{align}
where $\delta T_{h, H}$ are related to $\delta T_{1, 2}$ as,
\begin{align}
\left(\begin{array}{c}
\delta T_1 \\
\delta T_2  \\
\end{array}\right)
= R(\gamma) \left(\begin{array}{c}
\delta T_H \\
\delta T_h \\
\end{array}\right).
\end{align}

Mass counterterms are determined by imposing on-shell conditions to
renormalized two-point functions, Eqs.~(\ref{eq:pi_hh})-(\ref{eq:pi_HpHm}),
as follows
\begin{align}
  \hat{\Pi}_{\phi\phi}[m_\phi^2] =0.
\end{align}
The counterterms are then given by
\begin{align}
  \delta m_h^2 &= \frac{s_\gamma^2}{v}\delta T_1^{}
  -\frac{2s_\gamma c_\gamma}{v}\delta T_2^{} + \Pi_{hh}^{1\textrm{PI}}[m_h^2], \\[+5pt]
  \delta m_H^2 &= \frac{c_\gamma^2}{v}\delta T_1^{}
  +\frac{2s_\gamma c_\gamma}{v}\delta T_2^{} + \Pi_{HH}^{1\textrm{PI}}[m_H^2], \\[+5pt]
  \delta m_A^2 &= \Pi_{AA}^{1\textrm{PI}}[m_A^2], \\[+5pt]
  \delta m_{H^\pm}^2 &= \Pi_{H^+H^-}^{1\textrm{PI}}[m_{H^\pm}^2].
\end{align}

By imposing the following conditions;
\begin{align}
  &\frac{d}{dp^2}\hat{\Pi}_{\phi\phi}[p^2]\Big| _{p^2 = m_\phi^2} = 0, \,\,
  \frac{d}{dp^2}\hat{\Pi}_{G^0 G^0}[p^2]\Big| _{p^2 = m_{G^0}^2} = 0, \notag\\
  &\frac{d}{dp^2}\hat{\Pi}_{G^+ G^-}[p^2]\Big| _{p^2 = m_{G^\pm}^2} = 0,
\end{align}
wave function renormalization is fixed as
\begin{align}
  &\delta Z_\phi^{} =-\frac{d}{dp^2}\Pi_{\phi\phi}^\textrm{1PI}[p^2]\Big| _{p^2 = m_\phi^2},
  \\
  &\delta Z_{G0}^{} =-\frac{d}{dp^2}\Pi_{G^0 G^0}^\textrm{1PI}[p^2]\Big| _{p^2 = m_{G^0}^2},
  \\
&\delta Z_{G^\pm}^{} =-\frac{d}{dp^2}\Pi_{G^\pm G^\pm}^\textrm{1PI}[p^2]\Big| _{p^2 = m_{G^\pm}^2}.
\end{align}

We impose the following conditions to mixing two-point functions of renormalized fields,
\begin{align}
  \hat{\Pi}_{AG}^{}[m_A^2] = \hat{\Pi}_{AG}[0]=0, \,\,\,
  \hat{\Pi}_{HG}^{}[m_{H^\pm}^2] = \hat{\Pi}_{HG}[0]=0,
  \label{eq:mix_reno}
  \end{align}
such that mass eigenstates are diagonalized on mass shell.
This determines the renormalization conditions for
$\delta C_{AG}$, $\delta C_{GA}$, $\delta C_{HG}$ and $\delta C_{GH}$,
\begin{align}
  \delta C_{AG}&= \frac{1}{m_A^2}\left(
  \frac{\delta T_2}{v} + \Pi_{AG}^\textrm{1PI}[0]\right), \\
  \delta C_{GA}&=\frac{1}{m_A^2}\left(
  -\frac{\delta T_2}{v} - \Pi_{AG}^\textrm{1PI}[m_A^2]\right), \\
  \delta C_{HG}&=\frac{1}{m_{H^\pm}^2}\left(
  \frac{\delta T_2}{v} + \Pi_{HG}^\textrm{1PI}[0]\right), \\
  \delta C_{GH}&=\frac{1}{m_{H^\pm}^2}\left(
  -\frac{\delta T_2}{v} - \Pi_{HG}^\textrm{1PI}[m_{H^\pm}^2]\right).
\end{align}

For the CP-even states, as in the case of the CP-odd states,
we should impose on-shell condition on the two-point function,
\begin{align}
  \hat{\Pi}_{hH}^{}[m_h^2] = \hat{\Pi}_{hH}[m_H^2]=0,
\end{align}
which leads to the relations between $\delta \gamma$, $\delta C_{hH}$ and
$\delta C_{Hh}$:
\begin{align}
  \delta A&=\frac{1}{m_H^2 - m_h^2}
  \Big( -\frac{2s_\gamma c_\gamma}{v}\delta T_1
  +\frac{2c_{2\gamma}}{v}\delta T_2 \notag\\
  &+ \Pi_{hH}^{1\textrm{PI}}[m_h^2]
  +\Pi_{hH}^{1\textrm{PI}}[m_H^2]
  \Big), \label{eq:delA_delB1} \\
  \delta B&= \frac{1}{m_H^2 - m_h^2}
  \left( \Pi_{hH}^{1\textrm{PI}}[m_h^2] - \Pi_{hH}^{1\textrm{PI}}[m_H^2]
  \right), \label{eq:delA_delB2}
  \end{align}
where
\begin{align}
  \delta A &\equiv 2\delta \gamma -\delta C_{hH} + \delta C_{Hh}, \,\,\,
  \delta B \equiv \delta C_{hH} + \delta C_{Hh}.
  \label{eq:AandB}
\end{align}

In order to fix the three counterterms $\delta \gamma$, $\delta C_{hH}$ and $\delta C_{Hh}$,
an additional condition is required.
We employ a minimal subtraction renormalization condition to the three point functions,
which requires $\delta C_{hH}^{}$ to absorb only the divergent part of
the $HZZ$ vertex at one-loop level for $p_1^2=m_Z^2, p_2^2=m_Z^2, q^2=m_H^2$,
\begin{align}
  \hat{\Gamma}_{HZZ}^1[m_Z^2, m_Z^2, m_H^2]\Big|_\textrm{div. part} = 0,
 \label{eq:MS_con}
\end{align}
where $\hat{\Gamma}_{HZZ}^1$ is the scalar part of the $HZZ$ vertex function,
as defined through
\begin{align}
  \Gamma_{\varphi ZZ}^{\mu\nu} &= \Gamma_{\varphi ZZ}^1 g^{\mu\nu}
  + \Gamma_{\varphi ZZ}^2 \frac{p_1^\mu p_2^\nu }{m_Z^2}
  + i\Gamma_{\varphi ZZ}^3 \epsilon^{\mu\nu\rho\sigma}
  \frac{p_{1,\rho}^{}p_{2,\sigma}^{}}{m_Z^2}.
  \label{eq:form_factors}
\end{align}
By using Eqs.~(\ref{eq:delA_delB1}), (\ref{eq:delA_delB2}) and
the minimal subtraction condition given in Eq.~(\ref{eq:MS_con}),
we obtain explicit formulae for $\delta\gamma$, $\delta C_{hH}^{}$ and
$\delta C_{Hh}$,
\begin{align}
  \delta \gamma &= \frac{1}{2}(\delta A -\delta B + 2\delta C_{hH}^{}),\\
%  \end{align}
%
%\begin{align}
  \delta C_{hH}^{} &= -\frac{N_f^{C}s_\gamma}{32\pi^2 v^2}\Big\{
  c_\gamma(-2 m_f^2 + \rho_{ff}^2 v^2 + 2v^2 \rho_{ij}^{} \rho_{ji}^{}) \notag\\
  &- 2\sqrt{2}v s_\gamma^{} m_f^{} \rho_{ff}^{} \Big\}\Delta, \\
  \delta C_{Hh}^{} &=
  \frac{1}{m_H^2 - m_h^2}\left( \Pi_{hH}^{1\textrm{PI}}[m_h^2] -
  \Pi_{hH}^{1\textrm{PI}}[m_H^2] \right) - \delta C_{hH},
  \end{align}
where $\Delta \equiv 1/\epsilon -\gamma_E + \ln 4\pi +\ln \mu^2$.

%%%%%%%%%%%%%%%%%%%%%%%%%%%%%%%%%%%%%%
%%%%%%%%%%%%%%%%%%%%%%%%%%%%%%%%%%%%%%
\subsection{Renormalized vertices}\label{sec:reno_ver}
%%%%%

The renormalized scalar form factor of the $\phi ZZ$ vertex ($\phi = h, H$)
is composed of the tree level contribution, the counterterms,
and 1PI diagram contributions,
\begin{align}
  \hat{\Gamma}_{\phi ZZ}^1[p_1^2, p_2^2, q^2] =
  \Gamma_{\phi ZZ}^\textrm{tree} +
  \delta \Gamma_{\phi ZZ} +
  \Gamma_{\phi ZZ}^\textrm{1,1PI}[p_1^2, p_2^2, q^2].
  \end{align}
The counterterms are given by
\begin{align}
  \delta \Gamma_{hZZ}^1= &\frac{2m_Z^2}{v^2}  \sin(-\gamma)
  \left(\frac{\delta m_Z^2}{m_Z^2} - \frac{\delta v}{v}
  + \frac{1}{2}\delta Z_h^{} + \delta Z_Z\right) \notag\\
  &+ \frac{2m_Z^2}{v}  \cos\gamma \, \delta C_{Hh}^{},
 \label{eq:dG1_hZZ} \\
  \delta\Gamma_{HZZ}^1=&\frac{2m_Z^2}{v}  \cos\gamma
  \left(\frac{\delta m_Z^2}{m_Z^2} -\frac{\delta v}{v}
  +\frac{1}{2}\delta Z_H + \delta Z_Z  \right) \notag\\
  &+\frac{2m_Z^2}{v}  \sin(-\gamma) \, \delta C_{hH},
 \label{eq:dG1_HZZ}
\end{align}
where $\delta m_Z^2$, $\delta Z_Z^{}$ are the mass counterterm and
wave function renormalization of the $Z$ boson, respectively,
and their explicit formulae are given in Ref.~\cite{KKY_2HDM_full}.
We note that Eqs.~(\ref{eq:dG1_hZZ}) and (\ref{eq:dG1_HZZ})
have no $\delta\gamma$ dependence, the reason of which
can be traced to Eq.~(\ref{eq:Hh-shift}).

We here define the renormalized scaling factor of the $hZZ$ couplings in the following way;
\begin{align}
  \kappa_Z^{} = \frac{\Gamma_{hZZ}^{1}[(m_h+m_Z)^2,m_Z^2,m_h^2]}{\Gamma_{hZZ,\textrm{SM}}^1[(m_h + m_Z)^2,m_Z^2,m_h^2]},
  \label{eq:reno_kappa_Z}
\end{align}
where $\Gamma_{hZZ,SM}^1$ is the renormalized $hZZ$ coupling function in the SM.
We will numerically evaluate the deviation of $\kappa_Z^{}$ from 1
defined as $\Delta \kappa_Z^{} \equiv \kappa_Z^{} - 1$.

Before we enter numerical calculations,
in order to understand parameter dependence of $\Delta\kappa_Z^{}$,
we give an approximate formula for the one-loop corrected $hZZ$ coupling
that is effective in the decoupling limit,
i.e. the limit of $\varepsilon \ll 1$.
We further expand $\Delta\kappa_Z$ in $\varepsilon$:
\begin{align}
  \Delta\kappa_Z^{} &\simeq \; (\sin(-\gamma) -1 ) \notag\\
  -& \frac{1}{6}\frac{1}{16\pi^2}\sum_{\varphi=H,A,H^\pm}c_\varphi^{} \frac{m_\varphi^2}{v^2}
  \left(1- \frac{\mu_{22}^2}{m_\varphi^2}\right)^2 \notag\\
  +& \frac{\sqrt{2}N_t^C}{16\pi^2}\frac{m_t}{v}\rho_{tt}^{} \cos\gamma
  \sin^2\gamma
  \bigg\{(2 - \ln[m_H^2]) \notag\\
  -&2 \Big[B_0[m_h^2;m_t,m_t]-4m_t^2\frac{d}{dp^2}B_0[p^2;m_t,m_t]|_{p^2=m_h^2}
    \Big] \notag\\
  +& \;
   2\big[
  (v_f^2 + a_f^2)P_1
  -  (v_f^2 - a_f^2)P_2 \big]\bigg\} \notag\\
  &\simeq -\frac{\eta_6^2}{2} \,\varepsilon^2
  - \frac{1}{6}\frac{1}{16\pi^2}\sum_{\varphi=H,A,H^\pm} c_\varphi\frac{m_\varphi^2}{v^2}
  \left(1- \frac{\mu_{22}^2}{m_\varphi^2}\right)^2 \notag\\
  - & \frac{\sqrt{2}N_t^C}{16\pi^2}\frac{m_t}{v}\rho_{tt}^{} \eta_6^{} \, \varepsilon \,
  \bigg\{(2 - \ln[m_H^2]) \notag\\
  - &2 \Big[ B_0[m_h^2;m_t,m_t]-4m_t^2\frac{d}{dp^2}B_0[p^2;m_t,m_t]|_{p^2=m_h^2}
    \Big] \notag\\
  + & \;
   2\big[
     (v_f^2 + a_f^2)P_1  -  (v_f^2 - a_f^2)P_2\big]\bigg\}
   + \frac{1}{16\pi^2}\mathcal{O}(\varepsilon^2),
  \label{eq:dkz_appro}
  \end{align}
where $c_\varphi^{} = 2$ $(1)$ for $\varphi = H^\pm$ ($H, A$),
$N_t^C (=3)$ is the color factor of $t$, %i.e. $N_t^C = 3$.
$B_0$ is a Passarino-Veltman loop function [31],
and $P_{1, 2}^{}$ are combinations of various Passarino-Veltman loop functions,
defined as,
%where $B_0$ is a Passarino--Veltman loop function~\cite{PV_func},
%$c_\varphi^{} = 2$ (1) for $\varphi = H^\pm$ ($H, A$) and
\begin{align*}
P_1&\equiv  B_0[(m_h+ m_Z)^2;m_t,m_t] + B_0[m_Z^2;m_t,m_t] \notag\\
  &+ 2B_0[m_h^2;m_t,m_t]
+(4m_t^2 - m_h^2 -2 m_hm_Z) \notag\\
&\times C_0[(m_h + m_Z)^2, m_Z^2, m_h^2; m_t,m_t,m_t]
  \notag\\
  & - 8 C_{24}[(m_h + m_Z)^2, m_Z^2, m_h^2; m_t,m_t,m_t] \notag\\
  & \simeq -22.4, \\
  %%%%%
  P_2&\equiv
  B_0[(m_h + m_Z)^2;m_t,m_t] + B_0[m_Z^2;m_t,m_t] \notag\\
  +&(4m_t^2 - m_h^2)C_0[(m_h + m_Z)^2, m_Z^2, m_h^2; m_t,m_t,m_t] \notag\\
  & \simeq -22.6.
  %%%%
\end{align*}
%where $C_0$ and $C_{24}$ are also defined in Ref.~\cite{PV_func}.

%% We give here the reference values
%% \begin{align}
%%   &\textrm{Re}\,[P_1^{}[m_Z^2,(m_h+m_Z)^2,m_h^2;m_t]] \simeq -29.8, \\
%%   &\textrm{Re}\,[P_2^{}[m_Z^2,(m_h+m_Z)^2,m_h^2;m_t]] \simeq -40.5.
%%   \end{align}

The first, second and third terms in Eq.~(\ref{eq:dkz_appro}) correspond to
the tree level, extra scalar boson loop and fermion loop contributions, respectively.
Small $\varepsilon$ is the decoupling limit, which is a special case of alignment.

%%%%%
\begin{figure}[t]
 \centering
 \includegraphics[width=7.8cm]{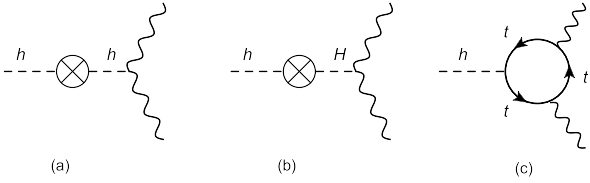}
 \vskip-0.3cm
 \caption{One loop diagrams contributing to the renormalized $hZZ$ vertex.}
 \label{fig:diag}
   \end{figure}
%%%%%

%%%%%%%%%%%%%%%%%%%%%%%%%%%%%%%%
\begin{figure*}[t]
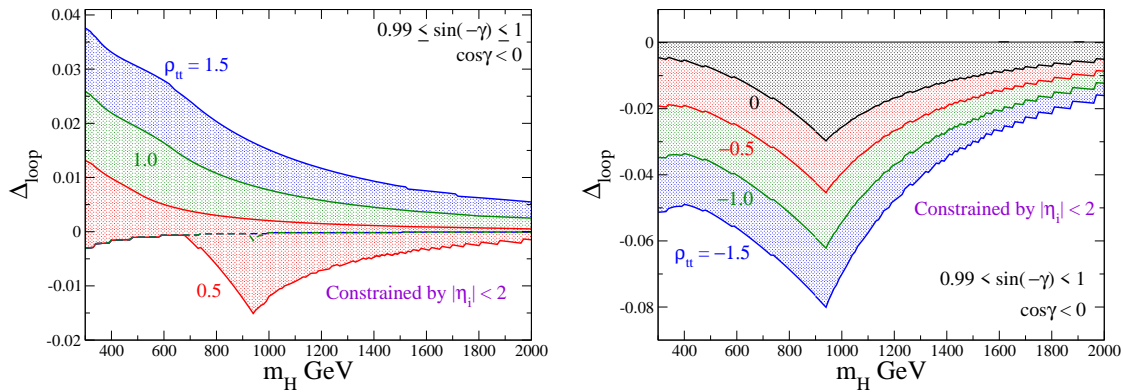

 \centering
 \includegraphics[width=7.2cm]{mH_maxmin_15.eps} \hspace{0.2cm}
 \includegraphics[width=7.2cm]{mH_maxmin_-15.eps} \hspace{0.2cm}
 \caption{$\Delta_{\rm loop}$ vs $m_H$ in the alignment limit,
   $0.99 \leq \sin(-\gamma) \leq 1$, and under perturbativity and vacuum stability constraints.
   The red, green and blue regions
   in the left (right) panel correspond to
$\rho_{tt}=0.5$, $1.0$, $1.5$ ($-0.5, \ -1.0,\ -1.5$), respectively. }
 \label{fig:mH_dsing}
\end{figure*}
%%%%%%%%%%%%%%%%%%%%%%%%%%%%%%%%

For the top quark loop contributions enclosed by $\{\ \}$ in Eq.~(\ref{eq:dkz_appro}),
the first, second  and third terms come from diagrams (a), (b) and (c)
in Fig.~\ref{fig:diag}, respectively.
%We see from Eq.~(\ref{eq:dkz_appro}) that
Besides $\rho_{tt}$ dependence, these dominant fermion loop contributions
have the mixing suppression factor $\cos\gamma$ (or $\varepsilon$).
Other contributions %to the $hZZ$ vertex
coming from fermion loops, such as $\rho_{ij}^f$ ($i\neq j$) contributions,
have square or higher power of $\cos\gamma$ ($\varepsilon$) suppression.
Thus, contributions from off-diagonal elements of the $\rho$ matrices are
subdominant in the alignment limit.
In addition,
the contributions from all kinds of fermion ($f'$) loops except the top quark
are suppressed by $m_{f'}/v$, so that they are also subdominant.

To facilitate our numerical study,
let us utilize Eq.~(\ref{eq:dkz_appro}) to discuss the radiative corrections to
$\kappa_Z$ and $\sin(-\gamma)$.
%From the definition of $\Delta \kappa_Z$,
The tree level contribution to the scaling factor $\kappa_Z$
is
\begin{align}
  \kappa_Z^\textrm{tree} = \sin(-\gamma),
 \label{eq:kap-tree}
\end{align}
which is keeping just the first term in Eq.~(\ref{eq:dkz_appro}),
and tree level means arising from the renormalized Higgs potential,
such that $\sin(-\gamma)$ is a renormalized quantity.
The other terms in Eq.~(\ref{eq:dkz_appro}) come from bosonic and fermionic loops,
hence we define the radiative shift due to loops
\begin{align}
  \Delta_{\rm loop} \equiv \kappa_Z^{} - \sin(-\gamma)
    = \Delta_{\rm loop}^{\rm bosonic} + \Delta_{\rm loop}^{\rho_{tt}},
  \label{eq:del_sin}
\end{align}
where $\kappa_Z^{}$ is the renormalized scaling factor
of Eq.~(\ref{eq:reno_kappa_Z}), and one can identify
the bosonic vs $\rho_{tt}$-induced top loop terms in Eq.~(\ref{eq:dkz_appro}).
That $\Delta_{\rm loop}$ contains both extra scalar boson loop
and $\rho_{tt}$-induced top loop contributions is a general result,
not just in the decoupling limit of Eq.~(\ref{eq:dkz_appro}).

Let us comment briefly on extra scalar boson loop contributions
to $\Delta_{\rm loop}$.
As we can see from Eq.~(\ref{eq:dkz_appro}),
the magnitude of the extra scalar loop correction strongly depends on the ratio of $\mu_{22}$ and $v$.
If $|\mu_{22}|$ is comparable to $v$,
the loop effect provides a quadratic power-like effect as $m_{\varphi}^2$.
%This arises from allowing Higgs couplings such as $\eta_3$, $\eta_4$ and $\eta_5$
%to have strength up to the perturbative bound of 2.
On the other hand, for $|\mu_{22}|^2 \gg v^2$,
the loop effect reduces as $1/m_\varphi^2$ according to the decoupling theorem.
Details of the non-decoupling effect of extra scalar loop corrections are explained in
Ref.~\cite{KKY_2HDM_full}.

Finally, it is useful to discuss the sign for each contribution.
The tree level and extra Higgs boson loop contributions decrease the $hZZ$ coupling
from the SM prediction.
However, for the top quark loop effect induced by $\rho_{tt}$,
whether the contribution attenuates or amplifies the value of the $hZZ$ coupling
depends on the sign of $\rho_{tt} \cos\gamma$.
If $\rho_{tt} \cos\gamma$ is negative,
the top quark loop contribution becomes the only one
that increases the $hZZ$ coupling as a main correction.
But if it is positive, it would further decrease the $hZZ$ coupling.

%%%%%%%%%%%%%%%%%%%%%%%%%%%%%%%%%%%%%%%%%%%%%%%%%%%%%%%%%%%%%%
\section{Numerical calculation}

In our numerical calculation, we take the following parameter values
as input~\cite{PDG2016}:
\begin{align}
  &m_Z^{} = 91.1876\ \textrm{GeV}, \,\,
  G_F^{} = 1.16638 \times 10^{-5}\ \textrm{GeV}^{-2}, \notag\\
  &\alpha_\textrm{EM}^{-1} = 137.035999, \,\,
  \Delta \alpha_\textrm{EM} = 0.06635, \notag\\
  &m_t^{} =173.34\ \textrm{GeV}, \,\,
  m_b^{} = 4.66\ \textrm{GeV}, \,\,
  m_c^{} =1.27\ \textrm{GeV}, \notag\\
  &m_\tau^{} =1.77686\ \textrm{GeV}, \,\,
  m_h^{} = 125\ \textrm{GeV}.
\end{align}
Although we should investigate effects from all kinds of $\rho_{f_i f_j}^{}$,
in this paper we concentrate on investigating contributions from $\rho_{tt}$
to the $hZZ$ vertex for simplicity.
Namely, we set $\rho_{ff}^{} = 0$ for $f = u, c, d, s, b, e, \mu, \tau$,
and $\rho_{f_i f_j}^{} = 0$ for $i\neq j$.
Contributions from the matrix components $\rho_{f_i f_j}$ ($i\neq j$) and $\rho_{ff}$
are subdominant, as mentioned at the end of Sec.~\ref{sec:reno_ver}.
In the following numerical calculation, we set $\rho_{tt}$ to be real for simplification of numerical calculation. 
We should also take into account the constraint from electroweak parameters "$S$, $T$, $U$"~\cite{Peskin_Takeuchi}. It is known that when mass differences of both extra neutral scalar bosons ($H, A$) and the charged scalar boson $H^\pm$ are too large, it conflicts with the data on T parameter~\cite{2HDM_STU}. Therefore, we hold $m_A = m_H = m_H^\pm$ in the following numerical calculation. In addition, too large a deviation from 1 of $\sin(- \gamma)$ conflicts with constraint from the electroweak parameters. However, since we consider only the case of $\sin(- \gamma)> 0.98$, parameter regions considered in this paper never conflicts with the constraints of the S, T, U parameter.

\begin{figure*}[ht]
  \centering
  \vspace{1cm}
 \includegraphics[width=8.2cm]{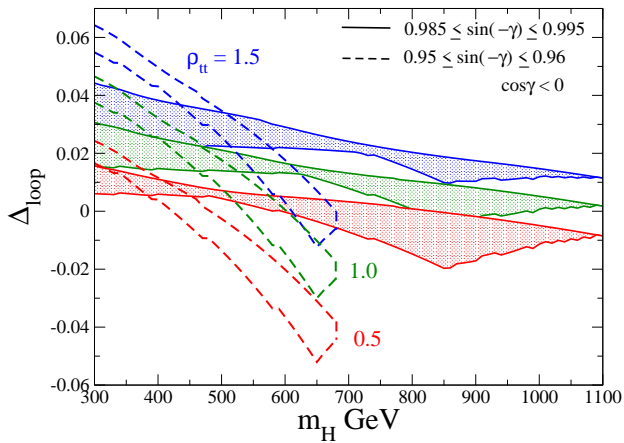} 
 \caption{$\Delta_{\rm loop}$ vs $m_H$ near alignment
  $0.985 \leq \sin(-\gamma) \leq 0.995$ (solid band) and
  $0.95 \leq \sin(-\gamma) \leq 0.96$ (dashed band),
  with settings the same as in Fig.~\ref{fig:mH_dsing}.}
 \label{fig:mH_dsing2}
\end{figure*}

%%%%%%%%%%%%%%%%%%%%%%%%%%%%%%%%
%%% radiative shift of sin\gamma
%%%%%%%%%%%%%%%%%%%%%%%%%%%%%%%%

We illustrate in Fig.~\ref{fig:mH_dsing} the range of variation for $\Delta_{\rm loop}$
by scanning $m_H$ and $\mu_{22}$ within the constraints of perturbativity and vacuum stability. 
We also scan $\sin(-\gamma)$, but limit the range to $0.99 \leq \sin(-\gamma) \leq 1$,
as we are interested in the $\rho_{tt}^{}$ effect in the alignment limit.
In the left (right) panel, the red, green and blue regions indicate the results for
$\rho_{tt}=0.5$, $1.0$, $1.5$ ($-0.5, \ -1.0,\ -1.5$), respectively.
Let us try to understand the features.

The peaking of $|\Delta_{\rm loop}|$ at $m_H \sim 950$~GeV
in the right panel of Fig.~\ref{fig:mH_dsing}
can be understood through the approximate formula, Eq.~(\ref{eq:dkz_appro}).
By the non-decoupling effect of the extra scalar bosons within
the perturbative bound, the strength of $|\Delta_{\rm loop}^{\rm bosonic}|$ ($\rho_{tt} = 0$ case)
can increase as $m_H^2$ for moderate $m_H$ values.
But when $m_H$ reaches 950 GeV and beyond, the perturbativity constraint
($\eta_i < 2$) cuts in, and large $m_H$ becomes dominated by large $|\mu_{22}|$,
hence $\Delta_{\rm loop}$ shrinks toward 0 in the decoupling limit of $m_H^2 \gg v^2$.
% begins to
Thus, the value of 950 GeV reflects our somewhat arbitrary choice
of perturbative bound, $|\eta_i| < 2$, for Higgs self-couplings.
As for the effect of $\rho_{tt}$, since $\rho_{tt} \cos\gamma > 0$,
a stronger $\rho_{tt}$ simply allows the negative
$\Delta_{\rm loop}^{\rm bosonic}$ effect to become even more negative.

%%%%%%%%%%%%%%%%%%%%%%%%%%%%%%%%
%%% Explanation for fig:easy
%%%%%%%%%%%%%%%%%%%%%%%%%%%%%%%%
\begin{figure*}[t]
 \centering
 \includegraphics[width=8cm]{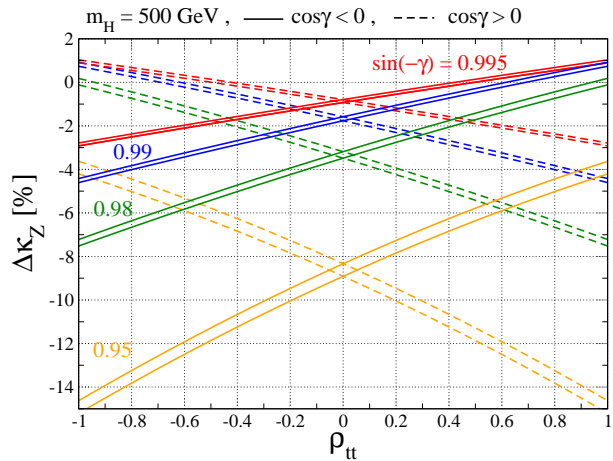} \hspace{0.5cm}
 \caption{$\Delta \kappa_Z$ vs $\rho_{tt}$ for $m_H^{} = 500$ GeV
 under perturbativity and vacuum stability constraints.
 For $\cos\gamma < 0$ ($\cos\gamma > 0$), solid (dashed)
 red, blue, green and orange bands are for
 $\sin(-\gamma)=0.995$, 0.99, 0.98 and 0.95, respectively.}
 \label{fig:easy}
\end{figure*}

More interesting is Fig.~\ref{fig:mH_dsing}(left),
where $\rho_{tt} \cos\gamma < 0$. For this case,
the $\rho_{tt}$ effect is opposite in sign to the bosonic loop contribution,
and moves $\Delta_{\rm loop}$ more positive.
For weak $\rho_{tt} = 0.5$, one sees similar peaking in negative values
for $\Delta_{\rm loop}$ as in Fig.~\ref{fig:mH_dsing}(right),
but for $m_H \lesssim 700$ GeV, one has $\Delta_{\rm loop} \gtrsim 0$
as $\rho_{tt}$ effect takes over.
For larger $\rho_{tt}$ values such as 1 or higher,
$\Delta_{\rm loop}$ is almost bound to be positive for the full $m_H$ range,
and can reach a few percent for low $m_H$ values.
For large $m_H$, decoupling again sets in, but more swiftly
than in Fig.~\ref{fig:mH_dsing}(right).
All these features reflect the fact that, for $\rho_{tt} \cos\gamma < 0$,
the $\rho_{tt}$ effect competes and cancels against the bosonic loop effect,
and $\Delta_{\rm loop} \sim 0$ is allowed, which means
$\kappa_Z$ could still have value $\kappa_Z^{\rm tree} = \sin(-\gamma)$.

The last statement brings about an interesting point, which we elucidate further.
The properties of the 125 GeV boson $h$ is in remarkable agreement
with the SM Higgs boson, and in the 2HDM context
this means we are close to alignment, i.e. $\cos\gamma \simeq 0$.
The alignment limit is usually understood in terms of the
decoupling limit of $m_H^2 \gg v^2$,
which makes extra Higgs boson search more difficult.
But could we have ``alignment without decoupling''~\cite{Gunion:2002zf, Pilaftsis, Craig:2013hca},
such that the exotic Higgs bosons are not so heavy, making them more amenable to search?
We find from our current study with potentially large $\rho_{tt}$
and sizable exotic Higgs couplings,
their effects could mutually cancel for $\rho_{tt} \cos\gamma < 0$,
such that alignment is indeed ``accidental'',
or Nature's design to keep the exotic Higgs doublet well hidden.

We plot $\Delta_{\rm loop}$ vs $m_H$ in Fig.~\ref{fig:mH_dsing2}
 for $0.985 \leq \sin(-\gamma) \leq 0.995$ (solid band)
 and $0.95 \leq \sin(-\gamma) \leq 0.96$ (dashed band), where
even $\sin(-\gamma) \simeq 0.955$ is still close to alignment,
with $|\cos\gamma| \simeq 0.3$.
The difference from Fig.~\ref{fig:mH_dsing} is that
$\sin(-\gamma) = 1$ is excluded, so $\cos\gamma$ cannot vanish.
One now sees the trend that, as $m_H$ increases,
$\Delta_{\rm loop}$ extends to more negative values, until
the bands are cut off by the perturbativity constraint.
For the less aligned case of $\sin(-\gamma) \sim 0.955$,
 the drop can be as much as $-0.07$, while
for the closer to aligned case of $\sin(-\gamma) \sim 0.98$,
 the drop is milder and can be of order $-0.04$.
The point is that we could have $\Delta_{\rm loop} \simeq 0$
and $\sin(-\gamma) \simeq 1$, but for moderate $m_H$ values
--- alignment without decoupling.
We note that, with $\sin(-\gamma)$ determined by the
renormalized Higgs potential, with parameters largely
not measured yet, we are far from knowing its true
value, except that alignment seems to hold to good extent.

With $\Delta_{\rm loop}$ better understood, we turn to
study numerically
\begin{align}
  \Delta \kappa_Z \equiv \kappa_Z - 1 = [\sin(-\gamma) - 1] + \Delta_{\rm loop},
  \label{eq:dkz}
\end{align}
the deviation of the $\kappa_Z$ observable of Eq.~(\ref{eq:reno_kappa_Z}) from 1.
First we reiterate that, e.g. for $\rho_{tt} = 1$ and for the case of
$0.985 \leq \sin(-\gamma) \leq 0.995$ in Fig.~\ref{fig:mH_dsing2},
one has $|\Delta \kappa_Z| \lesssim 0.01$, which is rather close to alignment limit,
but the full range of $m_H$ up to TeV is allowed.
We illustrate in Fig.~\ref{fig:easy}
the $\rho_{tt}^{}$ dependence of $\Delta\kappa_Z$ for $m_H^{} = 500$ GeV,
and for $\sin(-\gamma) = 0.995$, $0.99$, $0.98$ and $0.95$,
taking into account constraints from perturbativity
and vacuum stability on Higgs sector parameters.
For $\rho_{tt}^{} = 0$, the $hZZ$ coupling is affected
by the tree level mixing effect $\sin(-\gamma) - 1$,
and bosonic loop contributions $\Delta_{\rm loop} = \Delta_{\rm loop}^{\rm bosonic}$.
As discussed at the end of Sec.~\ref{sec:reno_ver},
these contributions reduce the value of the $hZZ$ coupling from SM~\cite{KKY_2HDM_full}.
For $\cos\gamma^{} < 0$, the top loop contributions with negative $\rho_{tt}$
reduce further the value of the $hZZ$ coupling.
%from the tree level mixing effect and bosonic loop contributions.
However, if $\rho_{tt}^{}$ is positive, the top loop effects
increase the value of the $hZZ$ coupling, i.e. it works against the bosonic contributions.
The value of $\Delta\kappa_Z^{}$ for $\sin(-\gamma) = 0.995$, $0.99$ and $0.98$
turns positive at $\rho_{tt}^{} \sim 0.5$, $0.7$ and $1$, respectively,
for $\cos\gamma < 0$.
For $\cos\gamma > 0$, the inclination of $\Delta\kappa_Z^{}$ is opposite to
the $\cos\gamma < 0$ case.

If the $hZZ$ coupling can be determined by experiment with some precision,
we can obtain the value of $\rho_{tt}$ for a given $\sin(-\gamma)$ value.
The combined LHC Run 1 data~\cite{LHC_Run1_Higgs} gives the 1$\sigma$ range of
$-6\% \leq \Delta\kappa_Z \leq 13 \%$ for the $hZZ$ coupling,
which is not yet discriminating enough to obtain information on
the value of $\rho_{tt}^{}$,
although it does disfavor $\sin(-\gamma) \lesssim 0.95$ for $\rho_{tt} \cos\gamma > 0$,
i.e. an expression for alignment.
%from the latest data of $\Delta\kappa_Z^{}$ because the measurement uncertainty is still large.
With full HL-LHC data, and at future colliders such as the the ILC
and the Compact LInear Collider (CLIC)~\cite{CLIC:2016zwp},
$\kappa_Z^{}$ is expected to be measured with higher accuracy as follows,
\begin{align}
\sigma(\kappa_Z^{}) &\simeq 2\% \quad\,\ \ \textrm{HL-LHC~\cite{HWGR}},
   \label{eq:kappa Z} \\
\sigma(\kappa_Z^{}) &\simeq 0.5\% \quad\, \textrm{ILC500~\cite{HWGR}},
   \label{eq:kappa Z ILC} \\
\sigma(\kappa_Z^{}) &\simeq 0.8\% \quad \textrm{CLIC350~\cite{CLIC}}.
   \label{eq:kappa Z CLIC}
\end{align}
Here ILC500 means the combination of $\sqrt{s}=250$ GeV run with
$L(\textrm{integrated luminosity}) = 250$ fb$^{-1}$
and $\sqrt{s}=500$ GeV with $L = 500$ fb$^{-1}$, while
CLIC350 is the staged CLIC~\cite{CLIC:2016zwp}
with $\sqrt{s}=350$ (and 380) GeV and $L = 500$ fb$^{-1}$.
With such precision obtainable in the future,
one could extract information on $\rho_{tt}^{}$ within uncertainties.
For example, for $m_H = 500$ GeV, if $\Delta \kappa_Z^{}$ is
measured at the central value of $-5\%$ at the HL-LHC (ILC500),
$|\rho_{tt}| \simeq - 0.74 \pm 0.87$ ($\pm 0.29$) and
$+ 0.42 \pm 1.16$ ($\pm 0.30$) are implied for
$\sin(-\gamma)=0.95$ and 0.98, respectively,
where errors reflect both %2$\sigma$
measurement %uncertainties
and theoretical uncertainties.
Therefore, indirect detection by $hZZ$ coupling measurements can
probe $|\rho_{tt}|$ for given value of $\sin(-\gamma)$,
while $B$ physics experiments can place only an upper bound.
We have also made clear the usefulness of an ILC, even if the energy
is below $H$ production threshold.

\begin{figure*}[t]
 \centering \hspace{-0.2cm}
\includegraphics[width=7cm]{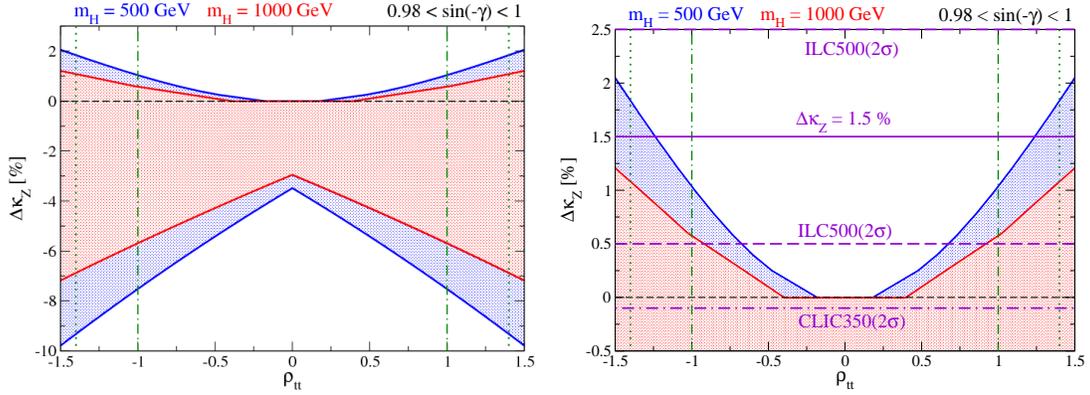} \hspace{0.1cm}
\includegraphics[width=7cm]{rhot_maxmin98_up3.eps}
 \caption{
   Range of $\Delta\kappa_Z$ for a given $\rho_{tt}$ value
   in the alignment limit, $0.98 \leq \sin(-\gamma) \leq 1$, and
   under perturbativity and vacuum stability constraints.
   Blue (red) region is for $m_H = 500$ (1000) GeV, with $\rho_{tt}$
   outside of dot-dashed (dotted) vertical lines excluded by $B_s$ mixing.
   In the right panel, which is a zoomed-in view for $\Delta \kappa_Z \gtrsim 0$,
   a horizontal solid line shows a favorable central value of $\Delta\kappa_Z$, with
   horizontal dashed (dot-dashed) lines indicating 2$\sigma$ error bars at the ILC500 (CLIC350). }
 \label{fig:rhott_dkz}
 \end{figure*}

It is difficult to compare the constraint from indirect search
with that from the direct search for $H$ studied in Ref.~\cite{carena_zhen},
i.e. heavy scalar search through $gg \to H/A \to t\bar{t}$ process
at HL-LHC, because the latter study corresponds to $\sin(-\gamma)=1$ in a 2HDM.
Let us compare the alignment limit (such as $\sin(-\gamma)=0.995$)
with the result of Ref.~\cite{carena_zhen}.
Suppose the measured central value is $\Delta\kappa_Z=0$ at ILC500.
In that case, as can be read from Fig.~\ref{fig:easy}, the 2$\sigma$ constraint from ILC is
$-0.1 < \rho_{tt} < 1$ for $\cos\gamma < 0$
($-1 < \rho_{tt} < 0.1$ for $\cos\gamma <0$).
%Fixing the sign to $\cos\gamma > 0$,
The $hZZ$ coupling precision measurement %may give a stronger bound than that from
would complement direct search bound at the LHC,
which gives $\rho_{tt} <0.5$~\cite{carena_zhen},
as it is hampered by complications from interference with $t\bar t$ background.
Our comparison, however, is based on rough estimates, and we
expect much progress by the time these measurements are made.

%%%%%%%%%%%%%%%%%%%%%%%%%%%%%%%%%%%%
%%%  explanation for rhott_dkz
%%%%%%%%%%%%%%%%%%%%%%%%%%%%%%%%%%%%

For a final perspective, we display in Fig.~\ref{fig:rhott_dkz}
the range of $\Delta\kappa_Z$ for a given value of $\rho_{tt}$,
for $m_H = 500$ GeV (blue shaded) and 1000 GeV (red shaded)
and close to alignment, $0.98 \leq \sin(-\gamma) \leq 1$.
We take into account perturbativity and vacuum stability bounds.
For $m_H = 500$ (1000) GeV, regions outside the dot-dashed (dotted) vertical lines
are excluded by $B_s$ mixing data.
The dependence of $\Delta\kappa_Z$ on $\rho_{tt}$ and $\sin(-\gamma)$
are as shown in Fig.~\ref{fig:easy}.
Thus, as $\Delta\kappa_Z^{}$ for a given value of $\rho_{tt}$ becomes more negative,
$\sin(-\gamma)$ deviates more from~1 (see Eq.~(\ref{eq:dkz})).
% We also can confirm numerically that
% the deviation in the $hZZ$ coupling tents to reduce because of the decoupling theorem
% as the extra Higgs bosons become to be heavier.

We see from Fig.~\ref{fig:rhott_dkz}(left) that, for $\sin(-\gamma) > 0.98$,
the most negative value for $\Delta \kappa_Z$ is about $-7.5 \%$
for $m_H=500$ GeV and $|\rho_{tt}^{}|=1$, with similar number
for $m_H = 1000 $ GeV and $|\rho_{tt}| = 1.5$.
Such reduction of $hZZ$ coupling can be uncovered by 
the HL-LHC (Eq.~(\ref{eq:kappa Z})), and would be quite interesting.
However, from Fig.~\ref{fig:easy} we see that, if $\sin(-\gamma)$ is smaller in value than 0.98,
%hence allowing larger $\cos\gamma$,
such negative values for $\Delta \kappa_Z$ can be realized by
non-decoupled bosonic loop effects for $\rho_{tt} = 0$.
%i.e. the exotic Higgs sector has relatively strong self-couplings.
%Namely, the minimum value of $\Delta\kappa_Z^{}$
%in each value of $\rho_{tt}$ depends on the value of the $\sin(-\gamma)$ and the criterion
%of perturbativity.
%
Without a clear handle on $\sin(-\gamma)$ (except that it is close to alignment),
which depends on many parameters, one cannot really determine $\rho_{tt}$.
Further measurements involving the exotic Higgs sector may help.
%
%On the other hand, as we mention in the explanation of Fig.~\ref{fig:easy},
%the maximum value does not change even if we take a different value
%as the criterion of the perturbativity bound and even if the value of $\sin(-\gamma)$ is
%smaller than 0.98.
%
The other direction, i.e. for $\Delta \kappa_Z >0$,
the situation is somewhat different.

We have commented that $\rho_{tt}$-induced top loop effects
would cancel against bosonic loop effects for $\rho_{tt}\cos\gamma < 0$,
which could give rise to alignment without decoupling, hence is of special interest.
In order to discuss the region where $\Delta\kappa_Z \gtrsim 0$,
as the possible range is narrower, we give a
zoomed-in view in Fig.~\ref{fig:rhott_dkz}(right).
Whether $m_H^{} = 500$ GeV or 1000 GeV,
in part because of the $B_s$ mixing constraint,
the $hZZ$ coupling can at most be $\sim 1 \%$ larger than
the SM prediction, which HL-LHC does not have the resolution to resolve
(although it can confirm a rather SM-like coupling,
further supporting alignment).

The $hZZ$ coupling, however, cannot be enhanced above SM
without the $\rho_{tt}$ effect of top loop diagrams.
Therefore, if such deviation is measured in future precision measurements
such as at the ILC500, it can probe the $\rho_{tt}$ coupling in the general 2HDM.
For example, suppose $\Delta \kappa_Z^{}$ is measured
with central value $+1.5\%$ at the ILC500 or CLIC350.
We mark this as a purple horizontal solid line in Fig.~\ref{fig:rhott_dkz}(right),
with dashed and dot-dashed horizontal lines indicating 2$\sigma$ error bars
at the ILC500 and CLIC350 (Eqs.~(\ref{eq:kappa Z ILC}) and (\ref{eq:kappa Z CLIC})), respectively.
In this case, $|\rho_{tt}| \lesssim 0.65$ ($0.9$) is excluded by $2 \sigma$
for $m_H = 500$ (1000) GeV by the ILC500, pointing towards an extra $\rho_{tt}$ Yukawa interaction.
Of course, if the central value falls at 1.0, then more data would be needed.
We remark that the comparison of CLIC350 with ILC500 is also an issue
of optimizing collision energy and run time.
If an evident deviation in the $hZZ$ coupling is not measured by the future precise measurement, we are hopeful for exploration for $\rho_{tt}$ by additional Higgs bosons searches using the signal $gg\to H/A \to t\bar{t}$ at the HL-LHC experiment~\cite{carena_zhen}.

%------------------------------------------------

%------------------------------------------------

%------------------------------------------------

\section{Conclusion}

We have calculated the renormalized $hZZ$ coupling at the one-loop level
by the on-shell and minimal subtraction scheme
in the general 2HDM without $Z_2 $ symmetry.
We numerically evaluated the one-loop corrected scaling factor of
the $hZZ$ coupling, in order to investigate the ability of
indirect detection of extra Yukawa interactions
with future Higgs boson coupling measurements.
In this paper, we focused on the top quark loop contributions
and heavy scalar boson loop contributions for simplicity.

By deriving an approximate formula for the renormalized scaling factor
$\kappa_Z$ of the $hZZ$ coupling, we make explicit that
the value of $\kappa_Z$ is determined by $\rho_{tt}^{}$,
the mass of extra scalar bosons $m_{\varphi}$, $\sin(-\gamma)$
and the sign of $\cos\gamma$.
Since $\kappa_Z$ would be $\sin(-\gamma)$ if one
considers only the renormalized Higgs potential,
we evaluate how much $\kappa_Z - \sin(-\gamma)$ is
shifted by radiative corrections in the alignment limit of $\sin(-\gamma) \simeq 1$.
We scan $m_H$ and $\mu_{22}$ keeping the assumption $m_H^{} = m_A^{} = m_{H^\pm}^{}$
under the constraints of perturbativity and vacuum stability
for some representative ranges for $\sin(-\gamma)$.
We find that the bosonic one-loop corrections always
shift $\kappa_Z - \sin(-\gamma)$ in the negative direction,
while the top loop correction induced by $\rho_{tt}$ depends on
the sign of $\rho_{tt}\cos\gamma$.
For $\rho_{tt}\cos\gamma > 0$, the $\rho_{tt}$ effect also
shifts $\kappa_Z - \sin(-\gamma)$ in the negative direction,
but for $\rho_{tt}\cos\gamma < 0$, the top loop effect
shifts $\kappa_Z - \sin(-\gamma)$ in the \emph{positive} direction,
and can cancel against the bosonic effect.
%increases in the positive (negative) direction as the value of $\rho_{tt}$ become large
%in the case for $\cos\gamma < 0$ ($\cos\gamma > 0$).
We have checked numerically that the magnitude of radiative shift
tends to vanish in the decoupling limit of $m_\varphi \to \infty$.

The cancellation effect mentioned above illustrates alignment without decoupling.
With $\kappa_Z - \sin(-\gamma)$ kept small by this cancellation,
even when both $|\rho_{tt}|$ and extra Higgs self-couplings are $O(1)$ or larger,
the observed ``alignment'' may be accidental,
and that exotic Higgs bosons could be around several hundred GeV in mass,
rather than the usual perception that alignment is realized by
the decoupling limit of very heavy exotic Higgs.
This makes the general 2HDM rather interesting.

%We investigated $\rho_{tt}$ dependence of $\kappa_Z$ with $m_H = 500$ GeV and 1000 GeV around the alignment limit as $0.98 \leq %\sin(-\gamma) \leq 1$.
%Although we obtained the lowest value of $\Delta\kappa_Z$
%under the range of $0.98 \leq \sin(-\gamma) \leq 1$ and the criterion of $|\eta_i < 2|$,
%the lowest value depends on the choice of the range and the perturbativity criterion value.
%On the other hand,
%the maximum value does not change even if we take a different value
%as the criterion of the perturbativity bound and even if the value of $\sin(-\gamma)$ is
%smaller than 0.98.
%
Future precision measurements such as at the ILC (and even the HL-LHC)
can survey $|\rho_{tt}|$ when $hZZ$ coupling is significantly lower than one,
for each value of $\sin\gamma$ and $m_\varphi^{}$,
while $B$ physics experiments and direct search of heavy scalar bosons at LHC
can place only upper bounds on $|\rho_{tt}|$.
However, given that bosonic corrections reduce the $hZZ$ coupling also,
if $\sin(-\gamma)$ is less than, say 0.98, one may not be able
to tell apart a purely bosonic effect, or that from $\rho_{tt}$.
But we have numerically showed that the $hZZ$ coupling
cannot be larger than the SM predicted value without the
$\rho_{tt}$-induced top quark loop effect, although the effect
is at the percent level.
If the $hZZ$ coupling turns out to be 1\% or more larger than the SM value,
the deviation can be sensed by the precision measurement at the ILC,
and would be definite evidence of the extra Yukawa interaction.
But the run time needed may exceed the definition of ILC500.
Of course, a higher energy ILC (or CLIC) could possibly discover the
exotic heavy Higgs bosons directly,
in this interesting case of alignment without decoupling.

Although we took into account the effect of extra Yukawa interaction
for only the top quark, other fermion loop effects arising from
extra Yukawa interactions should also be evaluated.
For example, the effect of $\rho_{cc}$ has not been explored much by $B$ physics
and LHC experiments.
Furthermore, we should investigate not only the effects of
the real part of $\rho_{ij}^{}$, which is what is studied in this paper
for simplicity, but we should also explore the impact of the imaginary part.
The imaginary parts, or CP phases of $\rho_{ij}$ could be of essential importance
for the generation of matter-antimatter asymmetry of the Universe.

%------------------------------------------------
\section*{acknowledgment}

We thank M. Kohda for discussions.
WSH is supported by grants MOST 104-2112-M-002-017-MY2,
MOST 105-2112-M-002-018 and NTU 105R8965, and
MK is supported by MOST 106-2811-M-002-010.

%------------------------------------------------
  \begin{appendix}
\section{1PI diagram contributions}

We give fermion loop contributions to the tadpoles, the two-point functions and
the three point functions
at the one-loop level by using Passarino-Veltman functions~\cite{PV_func}
whose notation is same as those in Ref.~\cite{Hagiwara}.
Explicit forms of 1PI bosonic loop contributions necessary for the renormalized $hZZ$ coupling are given in Ref.~\cite{KKY_2HDM_full}.

The 1PI tadpole diagrams for $h$, $H$ are calculated by
\begin{align}
  T_{h,F}^{1\textrm{PI}} &= \sum_{f}\frac{4N_f^C}{16\pi^2}\lambda_{hff}^{} m_f^{} A[m_f^{}],\\
  T_{H,F}^{1\textrm{PI}} &= \sum_{f}\frac{4N_f^C}{16\pi^2}\lambda_{Hff}^{} m_f^{} A[m_f^{}],
\end{align}
where $N_f^C$ indicates the color factor of $f$, and explicit formulae of $\lambda_{\phi ff}$
are given in Eqs.~(\ref{eq:lff}) and (\ref{eq:hff}).

The two-point function of $h$ and $h$--$H$ mixing
are corrected by the following 1PI diagrams,
%\onecolumn
%    [　　
\begin{align}
  &\Pi_{hh,F}^{1\textrm{PI}}[p^2]=
  - \frac{4N_f^C}{16\pi^2} \Bigg\{
  (\lambda_{hff}^{})^2A[m_f] \notag\\
  &  +\frac{1}{2}(\lambda_{hff}^{})^2 \Big[ 4m_f^2 -p^2\Big] B_0[p^2;m_f,m_f] \notag\\
  &  -\frac{1}{2}(\delta_f^h)^2 \Big[ 2A[m_f]  -p^2B_0[p^2;m_f,m_f] \Big]
   \Bigg\}
 \notag\\
 & - \frac{N_i^C \cos^2\gamma}{32\pi^2}\bigg\{
    2 \rho_{ij}^A m_i m_j B_0[p^2; m_i,m_j] \notag\\
 & \quad\quad\quad\ \, + \rho_{ij}^B \Big[ A[m_i] + A[m_j] \notag \\
 & \quad\quad\quad\quad\quad\ +(m_i^2 + m_j^2 - p^2)B_0[p^2; m_i,m_j] \Big]
 \bigg\}, \\
%\end{align}
%
%\begin{align}
  &\Pi_{hH,F}^{1\textrm{PI}}[p^2]=
  - \frac{4N_f^C}{16\pi^2} \Bigg\{
  \lambda_{hff}^{} \lambda_{Hff}^{}A[m_f] \notag\\
  & + \frac{1}{2}\lambda_{hff}^{} \lambda_{Hff}^{}
  \Big[ 2m_f^2 -p^2\Big] B_0[p^2;m_f,m_f] \notag\\
  &  -\frac{1}{2}\delta_f^h \delta_f^H \Big[ 2A[m_f]  -p^2B_0[p^2;m_f,m_f]\Big]
   \Bigg\}
 \notag\\
  %%%
 & - \frac{N_i^C}{32\pi^2}\sin\gamma \cos\gamma\bigg\{
 2 \rho_{ij}^A m_i m_j B_0[p^2; m_i,m_j] \notag\\
 & \quad\quad\quad\ \, +\rho_{ij}^B \Big[ A[m_i] + A[m_j] \notag \\
 & \quad\quad\quad\quad\quad\  +(m_i^2 + m_j^2 - p^2)B_0[p^2; m_i,m_j] \Big]
  \bigg\},
%% \
\end{align}
where ($i\neq j$)
\begin{align}
  \delta_f^h &= -\frac{\cos\gamma^{}}{2\sqrt{2}}\left(\rho_{ff}^{} - \rho_{ff}^\ast\right),
  \,\,\,\,
  (f=t, b, c, s, u, d)\\
  %%%
  \delta_f^H &= -\frac{\sin\gamma^{}}{2\sqrt{2}}\left(\rho_{ff}^{} - \rho_{ff}^\ast\right),\,\,\,\,
  (f=t, b, c, s, u, d) \\
  \rho_{ij}^A &= \rho_{ij} \rho_{ji} + \rho_{ij}^\ast \rho_{ji}^\ast, \\
  \rho_{ij}^B &= \rho_{ij} \rho_{ij}^\ast + \rho_{ji} \rho_{ji}^\ast.
  %% %%%
  %% \lambda_{ij}^h &=
  %% -\frac{c_\gamma^{}}{2\sqrt{2}}(\rho_{ij} + \rho_{ji}^\ast), \,\,\,\,
  %% (i\neq j), \\
  %% %%%
  %% \delta_{ij}^{h} &= -\frac{c_\gamma^{}}{2\sqrt{2}}(\rho_{ij} - \rho_{ji}^\ast),
  %% \,\,\,\,
  %% (i\neq j), \\
  %% %%%%
  %% \lambda_{ij}^H &=
  %% -\frac{s_\gamma^{}}{2\sqrt{2}}(\rho_{ij} + \rho_{ji}^\ast), \,\,\,\,
  %% (i\neq j), \\
  %% %%%
  %% \delta_{ij}^H &= -\frac{s_\gamma^{}}{2\sqrt{2}}(\rho_{ij}^{} - \rho_{ji}^\ast),
  %%   \,\,\,\,
  %% (i\neq j).
\end{align}

The 1PI diagram contributions to the $hZZ$ and $HZZ$ vertex form factors
defined in Eq.~(\ref{eq:form_factors}) are given by
\begin{align}
  &\Gamma_{\phi ZZ}^\textrm{1,1PI}(p_1^2,p_2^2,q^2)
  = - \sum_F \frac{8N_f^C m_f}{16\pi^2}\frac{m_Z^2}{v^2} \lambda_{\phi ff}^{}
  \times \notag\\
  &\bigg\{
    (v_f^2 + a_f^2)\Big[(3p_1^2 + p_1\cdot p_2)C_{11} +(3p_1\cdot p_2 + p_2^2)C_{12} \notag\\
  & \ + 2p_1^2 C_{21} + 2p_2^2 C_{22} + 4p_1\cdot p_2 C_{23} + 2(D-2)C_{24} \Big] \notag\\
  & \ - (v_f^2 + a_f^2)\Big[(p_1^2 + p_1\cdot p_2)C_{11} +(p_1\cdot p_2 + p_2^2)C_{12} \notag\\
  & \ + p_1^2 C_{21} + p_2^2 C_{22} + 2p_1\cdot p_2 C_{23} + (D-2)C_{24} \Big]
  \bigg\},
\end{align}
where $D = 4 - \epsilon /2$, $C_{ij} \equiv C_{ij}\,[p_1^2,p_2^2,q^2;m_t]$,
and $v_f = I_f - \sin^2\theta_W Q_f$ and $a_f = I_f$ are
the vector and axial vector coupling  coefficients of the $Z\bar{f}f$ vertex.

\end{appendix}

%----------------------------------------------------------------------------------------
%	REFERENCE LIST
%----------------------------------------------------------------------------------------

%----------------------------------------------------------------------------------------

\end{document}